\documentclass[twocolumn]{aastex61}

\usepackage{amsmath}
\usepackage{placeins}
\received{April 8, 2022}
\revised{June 16, 2022}
\accepted{\today}
\submitjournal{ApJ}

\shorttitle{Radio observations of SN\,2016gkg}
\shortauthors{Nayana et al.}

\begin{document}

\title{Radio evolution of a Type IIb supernova SN\,2016gkg}

\correspondingauthor{Nayana A.J.}
\email{nayana.aj@iiap.res.in}

\author[0000-0002-8070-5400]{Nayana A. J.}
\affil{Indian Institute of Astrophysics, II Block, Koramangala, Bangalore 560034, India.}
\author[0000-0002-0844-6563]{Poonam Chandra}
\affiliation{National Radio Astronomy Observatory, 520 Edgemont Road, Charlottesville VA 22903, USA}
\affiliation{National Centre for Radio Astrophysics, Tata Institute of Fundamental Research, PO Box 3, Pune, 411007, India}
\author{Anoop Krishna}
\affiliation{N.S.S College, Ottapalam, Palakkad 679103, India}
\author{G.C. Anupama}
\affiliation{Indian Institute of Astrophysics, II Block, Koramangala, Bangalore 560034, India.}

\begin{abstract}
We present extensive radio monitoring of a type\,IIb supernova (SN\,IIb), SN\,2016gkg during $t \sim$ 8$-$1429 days post explosion at frequencies $\nu \sim$ 0.33$-$25 GHz. The detailed radio light curves and spectra are broadly consistent with self-absorbed synchrotron emission due to the interaction of the SN shock with the circumstellar medium. The model underpredicts the flux densities at $t \sim$ 299 days post-explosion by a factor of 2, possibly indicating a density enhancement in the CSM due to a non-uniform mass-loss from the progenitor. Assuming a wind velocity $v_{\rm w} \sim$ 200 km\,s$^{-1}$, we estimate the mass-loss rate to be $\dot{M} \sim$ (2.2, 3.6, 3.8, 12.6, 3.7, and 5.0) $\times$ 10$^{-6}$ $M_{\odot}$\,yr$^{-1}$ during $\sim$ 8, 15, 25, 48, 87, and 115 years, respectively before the explosion. The shock wave from SN\,2016gkg is expanding from $R \sim$ 0.5 $\times$ 10$^{16}$ to 7 $\times$ 10$^{16}$ cm during $t \sim$ 24$-$492 days post-explosion indicating a shock deceleration index, $m$ $\sim$ 0.8 ($R \propto t^m$), and mean shock velocity $v \sim$ 0.1c. The radio data being inconsistent with free-free absorption model and higher shock velocities are in support of a relatively compact progenitor for SN\,2016gkg. 
\end{abstract}


\section{Introduction} \label{sec:intro}
Type IIb supernovae (SNe\,IIb) are a sub-class of core-collapse supernovae (CCSNe) characterized by the presence of broad HI absorption features in the early optical spectra. At later times, these HI lines disappear and He I feature becomes dominant in the spectra \citep{filippenko1997}, placing SNe\,IIb in between hydrogen-rich type II SNe and hydrogen poor type Ibc SNe. The progenitors of SNe\,IIb are understood to be stars that have lost most of their hydrogen envelope but not all. The hydrogen envelope could be lost either via radiatively driven winds \citep{smith2008} or via mass transfer by a binary companion \citep{yoon2010}.

Progenitor candidates have been identified for a few SNe\,IIb from pre-explosion images. A K-type supergiant in a binary system for SN\,1993J \citep{aldering1994,maund2004}, a yellow supergiant for SN\,2011dh \citep{arcavi2011,maund2011,vandyk2013,sahu2013}, and a massive star of mass $M \sim$ 20$-$25 $M_{\odot}$ for SN\,2008ax \citep{crockett2008,taubenberger2011}. Besides these direct detection efforts, the luminosity evolution of early light curve that maps the cooling envelope phase after the shock breakout can also put constraints on mass and radius of the progenitor star \citep{baron1993,swartz1993}.

Independent constraints on the progenitor properties can be obtained by studying the non-thermal radio emission from SNe\,IIb that arises as a result of the dynamical interaction between the supernova (SN) shock and the circumstellar medium \citep[CSM;][]{chevalier1982a,chevalier1998}. Radio observations uniquely probe the density structure of the CSM and thereby a longer period of mass-loss of the progenitor star \citep{weiler1986,chevalier1982a,chevalier1998}. Several SNe\,IIb exhibit luminous radio emission; a few examples are SN\,1993J \citep{weiler2007}, SN\,2001gd \citep{stockdale2003,stockdale2007}, SN\,2001ig \citep{ryder2004}, SN\,2003bg \citep{soderberg2006}, SN\,2011dh \citep{krauss2012,soderberg2012}, SN\,2011hs \citep{bufano2014}, and SN\,2013df \citep{kamble2016}.

\cite{chevalier2010} compiled a sample of radio bright SNe\,IIb and divided them based on their radio properties. The authors proposed two populations of SNe\,IIb; one with compact progenitors (SNe\,cIIb) and the other with extended progenitors (SNe\,eIIb). The SNe\,cIIb group shows faster shock velocities, less dense CSM, and compact progenitors in comparison with that of SNe\,eIIb. However, there exist a few examples \citep[e.g., SN\,2011dh][]{bersten2012,horesh2013,maeda2014} that suggest that the radio properties may or may not be a good indicator of the progenitor size. The progenitors of SNe\,IIb could be a continuum of objects between compact and extended stars instead of a sharp split like SNe\,eIIb and SNe\,cIIb.

This paper presents the radio follow-up observations of a type IIb supernova (SN\,IIb) SN\,2016gkg from $t \sim$ 8 to 1429 days over a frequency range of 0.3$-$25 GHz. The data include our observations taken with the Giant Metrewave Radio Telescope (GMRT) and the archival data from the Jansky Very Large Array (JVLA). We model the radio emission to investigate the mass-loss history of the progenitor system, the evolution of SN shock radius \& magnetic field, and irregularities in the CSM density.

The paper is organized as follows. In \S \ref{sec:sn2016gkg}, we present the compilation of various results on SN\,2016gkg from the literature. The details of observations and data reduction are presented in \S \ref{sec:obs}. We discuss the radio emission model and derive various parameters of the progenitor and environments in \S \ref{sec:radio-model} and \S \ref{sec:non-uniform csm}. The results are discussed in \S \ref{sec:discussion} and conclusions are drawn in \S \ref{sec:conclusion}.

\section{SN\,2016gkg}
\label{sec:sn2016gkg}
 SN 2016gkg was discovered by Buso \& Otero on 2016 Sep 20.18 (UT)\footnote{http://ooruri.kusastro.kyoto-u.ac.jp/mailarchive/vsnet-alert/20188} at a position $\alpha_{\rm J2000}$ = 01$^{\rm h}$34$^{\rm m}$14.40$^{\rm s}$, $\delta_{\rm J2000}$ = $-$29$^{\circ}$26$^{\prime}$24.20$^{\prime \prime}$. The SN is located at a distance of 26.4 $\pm$ 5.3 Mpc in the galaxy NGC\,613 \citep{nasonova2011}. The SN was classified as a SN\,IIb based on the optical spectroscopic observations \citep{tartaglia2017}. \cite{kilpatrick2017} model the early time optical light curve of SN\,2016gkg and derive the date of explosion to be $t_{\rm 0} =$ 2016 September 20.15$^{+0.08}_{-0.10}$ UT. We adopt this as the date of explosion through out this paper and all epochs $t$ mentioned are with respect to $t_{\rm 0}$.

SN\,2016gkg was extensively followed in the optical bands soon after the discovery which provided excellent coverage of its early evolution. The SN showed double peak structure in the optical light curve, the early peak due to the shock cooling of the hydrogen envelope of the progenitor star, and the later peak powered by radioactive decay \citep{bersten2018,kilpatrick2017,tartaglia2017}. Pre-explosion images of the field containing SN\,2016gkg taken in 2001 with the Hubble Space Telescope (HST) Wide Field Planetary-Camera 2 (WFPC2) is available in the archive \citep{bersten2018}. 

Various groups \citep{kilpatrick2017,tartaglia2017,piro2017,arcavi2017,bersten2018} attempted to constrain the properties of the progenitor of SN\,2016gkg by modeling the early peak of the light curve by shock cooling models. The progenitor radius estimates from these studies span a wide range $R_{*}\sim$ 40 $-$ 646 $R_{\odot}$ depending on the model and assumed structure of the hydrogen envelope of the progenitor.
\cite{piro2017} investigated the early peak of the optical light curve by numerically exploding a large number of extended envelope models and constrain the radius to be $R_{*} \sim$ 180--260 $R_{\odot}$. \cite{arcavi2017} fit the observed light curve with analytical shock cooling models \citep{nakar2014,piro2015,sapir2017} and estimate the progenitor radius to be $R_{*} \sim$ 40--150 $R_{\odot}$. \cite{bersten2018} modeled the cooling peak and estimated the radius of the hydrogen envelope to be $\sim$ 320 $R_{\odot}$. Modelling the initial rapid rise of the light curve using \cite{rabinak2011} model, \cite{kilpatrick2017} constrained the progenitor radius to be $R_{*} =$ 257$^{+389}_{-189}$. \cite{tartaglia2017} modeled the temperature evolution of initial peak of the light curve and estimated the progenitor radius to be $R_{*} \sim$ 48$-$124 $R_{\odot}$.

\cite{kilpatrick2017} detected a progenitor candidate of SN\,2016gkg in the archival HST image and estimated the luminosity and radius to be log(L/L$_{\odot}$) = 5.15 and $R_{*} =$138$^{+131}_{-103}$ $R_{\odot}$, respectively. The authors found that single star stellar evolution models fail to reproduce the derived progenitor properties whereas binary evolutionary tracks could reproduce them. \cite{tartaglia2017} identified two plausible progenitor candidates from HST imaging analysis and suggested a range of progenitor mass 15$-$20 $M_{\odot}$ and radius (150$-$320) $R_{\odot}$. \cite{kilpatrick2021} present post-explosion late-time ($t \sim$ 652$-$1795 days) HST observations of SN\,2016gkg and their improved astrometric allignment between the SN and progenitor candidate allowed them to constrain the progenitor to be a compact yellow super-giant of radius $\sim$ 70 $R_{\odot}$ with effective temperature ($T_{\rm eff}$) $\sim$ 10800 K. Late time ($t \sim$ 300--800 days) spectroscopic observations of SN\,2016gkg reveal multi component emission lines indicating the presence of material with different velocities possibly indicating an asymmetric explosion \citep{kuncarayakti2020}.

\section{Observations and Data Reduction}
\label{sec:obs}
\subsection{GMRT observations}
\label{subsec:gmrt-obs}
We carried out regular monitoring of SN\,2016gkg with the Giant Metrewave Radio Telescope (GMRT) during $t \sim$ 51 $-$ 1429 days at 0.33, 0.61, and 1.39 GHz. The data were recorded with an integration time of 16 seconds in full polar mode with a bandwidth of 33 MHz, split into 256 channels. 3C\,286 and 3C\,147 were used as the flux density calibrators and J0240$-$2309 was used as the phase calibrator. 

The GMRT data were inspected and calibrated using the Astronomical Image Processing Software \citep[AIPS; ][]{greisen2003} following standard procedure \cite[see ][]{Nayana2017}. The calibrated visibilities were imaged excluding the short $uv$ data to minimize the contribution from the extended host galaxy emission. The host galaxy is of flux density $\sim$ 178.3 mJy at 1.4 GHz with an angular size $\sim$ 48.5 $\times$ 30.4 arcsec$^{2}$ in the NVSS map \citep{condon1998}. The flux density of the SN was determined by fitting a Gaussian at the source position using task JMFIT\footnote{http://www.aips.nrao.edu/cgi-bin/ZXHLP2.PL?JMFIT}. The SN is $\sim$ 92 arcsec away from the center of the host galaxy, well beyond the continuum emission from the galaxy. However, we fit a zero-level baseline while fitting the Gaussian to account for any residual emission from the host galaxy. The details of GMRT observations are given in Table \ref{tab:gmrt}. The errors on flux densities quoted in Table \ref{tab:gmrt} are the sum of map rms values and a 10\% calibration uncertainty added in quadrature. The upper limits on radio flux densities are three times the map rms at the supernova position. The GMRT light curves are shown in Figure \ref{fig:lc-gmrt}.

\subsection{JVLA observations}
\label{subsec:JVLA observations}
The Karl G. Jansky Very Large Array (JVLA) observed SN\,2016gkg during $t \sim$ 8 -- 492 days covering a frequency range 2--25 GHz (archival data; PI: Maria Drout). The observations were done in standard continuum mode with a bandwidth of 2.048 GHz split into 16 spectral windows each of 128 MHz. 3C\,147 was used as the flux density calibrator, and J0145$-$2733 was used as the phase calibrator. 

\begin{figure*}
\begin{centering}
\includegraphics[scale=0.7]{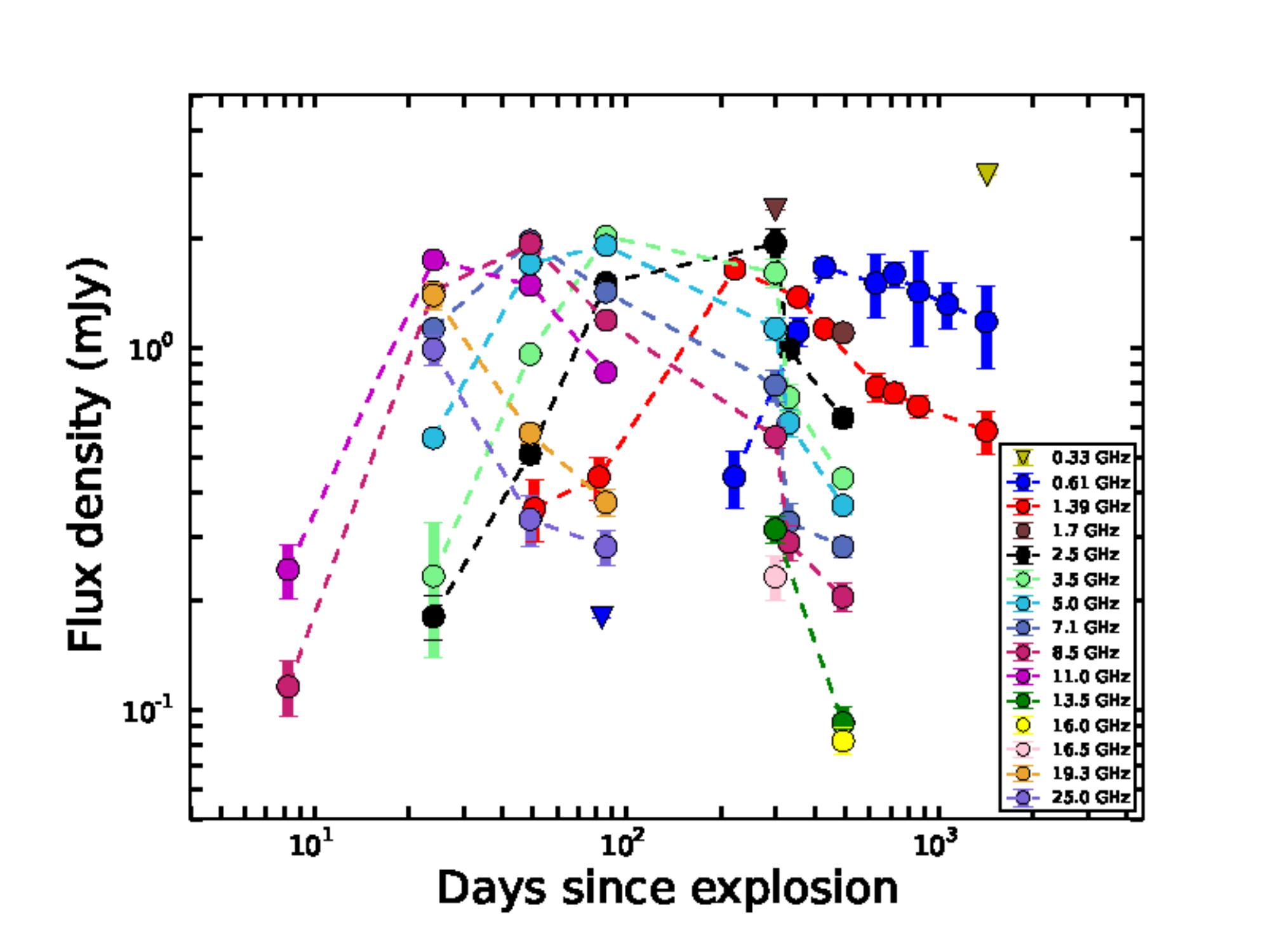}
 \caption{ \scriptsize{The GMRT and JVLA light curves of SN\,2016gkg at frequencies 0.33--25 GHz. The inverted triangle denotes 3$\sigma$ upper limits. The age of the SN is calculated assuming the date of explosion to be 2016 Sep 20.15 UT \citep{kilpatrick2017}.}}
    \label{fig:lc-gmrt}
    \end{centering}
\end{figure*}

The JVLA archival data were reduced using standard packages in Common Astronomy Software Applications \citep[CASA;][]{mcmullin2007}. We use the CASA task TCLEAN for imaging and we exclude short baselines to minimize the host galaxy emission while imaging. The flux density of the source is estimated by fitting a Gaussian at the source position using task IMFIT. The details of JVLA observations are given in Table\ \ref{tab:vla}. The errors on flux densities quoted in Table \ref{tab:gmrt} are the sum of map rms values and 10\% calibration uncertainties added in quadrature. Fig \ref{fig:lc-gmrt} shows the flux density evolution of SN\,2016gkg at frequencies 2$-$25 GHz.

\begin{deluxetable}{lccccc}
\tablecaption{Details of GMRT observations of SN\,2016gkg  \label{tab:gmrt}}
\tablehead{
\colhead{Date of Observation} & \colhead{Age\tablenotemark{a}} & \colhead{Frequency} & \colhead{Flux density} \\
\colhead{(UT)} & \colhead{(Day)} & \colhead{(GHz)} &  \colhead{(mJy)\tablenotemark{b}}  
}
\startdata
2016 Nov 09.76    &    50.61       &    1.39     &  0.36   $\pm$ 0.08  \\
2016 Dec 10.64    &    81.49       &    1.39     &  0.44   $\pm$ 0.07  \\
2017 Apr 30.16    &    222.01      &    1.39     &  1.65   $\pm$ 0.18   \\
2017 Sep 08.83    &    353.68      &    1.39     &  1.38   $\pm$ 0.15  \\
2017 Nov 23.63    &    429.48      &    1.39     &  1.13   $\pm$ 0.12  \\
2018 Jun 12.98    &    630.83      &    1.39     &  0.78   $\pm$ 0.10  \\ 
2018 Sep 08.85    &    718.70      &    1.39     &  0.75   $\pm$ 0.09  \\
2019 Jan 27.45    &    859.30      &    1.39     &  0.69   $\pm$ 0.09  \\
2020 Aug 07.09    &    1416.94     &      1.39   &  0.59   $\pm$ 0.10  \\
\hline 
2016 Dec 12.57    &    83.42       &    0.61    &  $<$0.18           \\
2017 Apr 29.29    &    221.14      &    0.61    &  0.44 $\pm$ 0.09  \\
2017 Sep 08.83    &    353.68      &    0.61    &  1.11 $\pm$ 0.15  \\
2017 Nov 23.63    &    429.48      &    0.61    &  1.67 $\pm$ 0.20  \\
2018 Jun 10.85    &    628.70      &    0.61    &  1.51 $\pm$ 0.34   \\
2018 Sep 09.83    &    719.68      &    0.61    &  1.60 $\pm$ 0.21   \\
2019 Jan 28.41    &    860.26      &    0.61    &  1.43 $\pm$ 0.44   \\
2019 Aug 20.92    &    1064.77     &    0.61    &  1.32 $\pm$ 0.23   \\
2020 Aug 05.05    &    1414.90     &    0.61    &  1.18 $\pm$ 0.32   \\
\hline
2020 Aug 18.82    &    1428.67     &   0.325    &     $\leq$3     \\
\enddata
\tablenotetext{a}{The age is calculated assuming 2016 Sep 20.15 (UT) as the date of explosion \citep{kilpatrick2017}.}
\tablenotetext{b}{The errors on the flux densities are the sum of error from task JMFIT and a 10\% calibration uncertainity added in quadrature.}
\end{deluxetable}

\begin{deluxetable}{lccccc}
\centering
\tablecaption{Details of JVLA observations of SN\,2016gkg  \label{tab:vla}}
\tablecolumns{6}
\tablenum{2}
\tablewidth{0pt}
\tablehead{
\colhead{Date of } & \colhead{Age\tablenotemark{a}} & \colhead{Frequency} & \colhead{VLA} & \colhead{Flux density} \\
\colhead{Observation (UT)} & \colhead{(Day)} &
\colhead{(GHz)} & \colhead{Array} & \colhead{(mJy)\tablenotemark{b}} 
}
\startdata
2016 Sep 28.39  & 8.24   & 8.549    & A  & 0.116 $\pm$ 0.023 \\
		        & 8.24   & 10.999   & A  & 0.244 $\pm$ 0.049 \\	
2016 Oct 14.21  & 24.06  & 2.499    & A  & 0.181 $\pm$ 0.031  \\
                & 24.06  & 3.499    & A  & 0.234 $\pm$ 0.097  \\
		        & 24.06  & 4.999    & A  & 0.564 $\pm$ 0.063  \\
		        & 24.06  & 7.099    & A  & 1.128 $\pm$ 0.115  \\
		        & 24.06  & 8.549    & A  & 1.430 $\pm$ 0.151  \\
		        & 24.06  & 10.999   & A  & 1.749 $\pm$ 0.184  \\
		        & 24.06  & 19.299   & A  & 1.396 $\pm$ 0.184  \\
		        & 24.06  & 24.999   & A  & 0.991 $\pm$ 0.139  \\
2016 Nov 08.21  & 49.06  & 2.499    & A  & 0.509 $\pm$  0.059   \\
                & 49.06  & 3.499    & A  & 0.960 $\pm$  0.104   \\	
	            & 49.06  & 4.999    & A  & 1.708 $\pm$  0.174  \\
	            & 49.06  & 7.099    & A  & 1.976 $\pm$  0.201  \\
	            & 49.06  & 8.549    & A  & 1.939 $\pm$  0.197  \\
	            & 49.06  & 10.999   & A  & 1.489 $\pm$  0.154  \\
	            & 49.06  & 19.299   & A  & 0.581 $\pm$  0.068  \\
	            & 49.06  & 24.999   & A  & 0.336 $\pm$  0.063  \\   
2016 Dec 15.01  & 85.86  & 2.499    & A  & 1.510 $\pm$  0.033  \\
                & 85.86  & 3.499    & A  & 2.031 $\pm$  0.208  \\	            
	            & 85.86  & 4.999    & A  & 1.917 $\pm$  0.193  \\
	            & 85.86  & 7.099    & A  & 1.427 $\pm$  0.144  \\
	            & 85.86  & 8.549    & A  & 1.192 $\pm$  0.123  \\
	            & 85.86  & 10.999   & A  & 0.856 $\pm$  0.089  \\
	            & 85.86  & 19.299   & A  & 0.374 $\pm$  0.050  \\
	            & 85.86  & 24.999   & A  & 0.282 $\pm$  0.043  \\ 
2017 Mar 19.84  & 299.08 & 1.749    & D & $<$ 2.4  &  -       \\
                & 299.08 & 2.499    & D & 1.940 $\pm$ 0.272    \\  
                & 299.08 & 3.499    & D & 1.612 $\pm$ 0.220    \\  
                & 299.08 & 4.999    & D & 1.131 $\pm$ 0.140    \\  
                & 299.08 & 7.099    & D & 0.788 $\pm$ 0.110    \\  
                & 299.08 & 8.549    & D & 0.567 $\pm$ 0.067    \\   
                & 299.08 & 9.499    & D & 0.485 $\pm$ 0.064    \\  
                & 299.08 & 13.499   & D & 0.315 $\pm$ 0.042    \\  
                & 299.08 & 16.499   & D & 0.233 $\pm$ 0.040    \\                 
\enddata
\tablenotetext{a}{The age is calculated using 2016 Sep 20.15 (UT) as the date of explosion \citep{kilpatrick2017}.}
\tablenotetext{b}{The errors on the flux densities are the sum of error from task IMFIT and a 10\% calibration uncertainity added in quadrature.}
\end{deluxetable}

\begin{deluxetable}{lccccc}
\centering
\tablecaption{Details of JVLA observations of SN\,2016gkg (continued) \label{tab:vla}}
\tablecolumns{6}
\tablenum{2}
\tablewidth{0pt}
\tablehead{
\colhead{Date of } & \colhead{Age\tablenotemark{a}} & \colhead{Frequency} & \colhead{VLA} & \colhead{Flux density}  \\
\colhead{Observation (UT)} & \colhead{(Day)} &
\colhead{(GHz)} & \colhead{Array} & \colhead{(mJy)\tablenotemark{b}} 
}
\startdata
2017 Aug 16.44  & 330.29 & 2.499    & C & 0.992 $\pm$ 0.108  \\
                & 330.29 & 3.499    & C & 0.729 $\pm$ 0.094  \\
                & 330.29 & 4.999    & C & 0.620 $\pm$ 0.082  \\
                & 330.29 & 7.099    & C & 0.330 $\pm$ 0.054  \\
                & 330.29 & 8.549    & C & 0.290 $\pm$ 0.044  \\
                & 330.29 & 9.499    & C & 0.270 $\pm$ 0.035 \\  
2018 Jan 25.04  & 491.89   & 2.499  & B  & 0.639 $\pm$ 0.075  \\   
                & 491.89   & 3.499  & B  & 0.436 $\pm$ 0.050  \\
                & 491.89   & 4.999  & B  & 0.367 $\pm$ 0.042  \\
                & 491.89   & 7.099  & B  & 0.282 $\pm$ 0.033  \\
                & 491.89   & 8.549  & B  & 0.205 $\pm$ 0.027  \\
                & 491.89   & 9.499  & B  & 0.200 $\pm$ 0.024  \\               
2018 Jan 26.04  & 492.89   & 1.749  & B  & 1.100 $\pm$ 0.112   \\
                & 492.89   & 13.499 & B  & 0.092 $\pm$ 0.014   \\
                & 492.89   & 15.999 & B  & 0.082 $\pm$ 0.011  \\
\enddata
\tablenotetext{a}{The age is calculated using 2016 Sep 20.15 (UT) as the date of explosion \citep{kilpatrick2017}.}
\tablenotetext{b}{The errors on the flux densities are the sum of error from task IMFIT and a 10\% calibration uncertainity added in quadrature.}
\end{deluxetable}


\section{Radio emission Model}
\label{sec:radio-model}
The general properties of radio emission from CCSNe have been discussed in detail by \cite{chevalier1982a,chevalier1998,weiler1986,weiler2002}. The radio light curves and spectra can be described using the ``mini-shell" model \cite[standard model;][]{chevalier1982a}. According to this model, the forward shock from the SN interacts with the ionized CSM established due to the stellar winds of the progenitor star. At the shock, particles are accelerated to relativistic velocities in amplified magnetic fields and emit synchrotron radiation. A fraction of post-shock energy density is distributed into magnetic fields ($\epsilon_{\rm B}$) and relativistic electrons ($\epsilon_{\rm e}$), and this fraction is assumed to be constant throughout the evolution of the ejecta. The observed radio light curves/spectra will be characterized by synchrotron radiation, where the low-frequency emission is significantly suppressed by an absorption component. The absorption can be free-free absorption (FFA) due to the ionized wind material along the line of sight \citep{weiler1986} or due to the same relativistic electrons that generate radio emission \citep[synchrotron self-absorption; SSA][]{chevalier1998}. The radio flux density initially rises rapidly and then declines as a result of the combined effects of non-thermal synchrotron emission and various absorption processes. 

This standard model of hydrodynamic evolution of ejecta assumes self-similar evolution of physical parameters across the shock \citep{chevalier1996}. The shock radius evolves as $R \sim$ $t^{m}$ where $m = (n-3)/(n-s)$, and $n$ indicates the outer density profile of the SN ejecta ($\rho_{\rm ejecta} \propto r^{-n}$). $s$ denotes the density profile of CSM ($\rho_{\rm CSM} \propto r^{-s}$) and the value of $s =2$ for a wind stratified medium.

We adopt the model from \cite{weiler1986} in a scenario where the dominant absorption process is FFA and follow a similar method as discussed in \cite{Nayana2018} and \cite{Nayana2020}. 
In an FFA model, the spectral and temporal evolution of radio flux densities $F(\nu,t)$ can be described as;

\begin{equation}
\label{eqn:ffa-general-flux}
F(\nu,t) = K_{1}\left(\frac{\nu}{5\, \rm GHz}\right)^{\alpha}\left(\frac{t}{10\, \rm days}\right)^{\beta}e^{-\tau_{\rm FFA}}
\end{equation}

\begin{equation}
\label{eqn:ffa-optical-depth}
\tau_{\rm FFA} = K_{2}\left(\frac{\nu}{5\, \rm GHz}\right)^{-2.1}\left(\frac{t}{10\, \rm days}\right)^{\delta}
\end{equation}

The multi-frequency radio flux density evolution in the case of a dominant SSA scenario can be modeled as \citep{chevalier1998}
\begin{equation}
\label{eqn:ssa-general-flux}
F(\nu,t) = K_{1}\left(\frac{\nu}{5\, \rm GHz}\right)^{2.5} \left(\frac{t}{10\, \rm days}\right)^{a} \left[1-e^{-\tau_{\rm SSA}}\right]
\end{equation}

\begin{equation}
\label{eqn:ssa-optical-depth}
\tau_{\rm SSA} = K_{2}\left(\frac{\nu}{5\, \rm GHz}\right)^{-(p+4)/2}\left(\frac{t}{10\, \rm days}\right)^{-(a+b)}
\end{equation}

In the above equations, $K_1$ and $K_2$ denote the flux density and optical depth at 5 GHz on $t =$ 10 days post-explosion, respectively. In equations \ref{eqn:ffa-general-flux} and \ref{eqn:ffa-optical-depth}, $\alpha$ represents the spectral index ($F_{\nu} \propto \nu^{\alpha}$) and $\beta$ represents the temporal index of radio flux densities. The term exp(${-\tau_{\rm FFA}}$) corresponds to the attenuation due to the absorption by a unifrormly distributed ionized CSM external to the radio emitting region. '$\delta$' denotes the temporal evolution of $\tau_{\rm FFA}$, and is related to $\alpha$ and $\beta$ as $\delta = \alpha-\beta-3$. Assuming the CSM is created due to a steady stellar wind ($\rho_{\rm CSM} \propto$ r$^{-2}$), the shock decceleration parameter; $m$ can be written as $m =-\delta$/3 where  $m=(n-3)/(n-2)$. 

In equations \ref{eqn:ssa-general-flux} and \ref{eqn:ssa-optical-depth}, $a$ and $b$ denote the temporal indices of flux densities in the optically thick ($F \propto t^a$) and thin phase ($F \propto t^{-b}$), respectively. $\tau_{\rm SSA}$ is the optical depth due to SSA and $p$ is the electron energy power-law index ($N(E) \propto E^{-p}$) which is related to $\alpha$ as $p = 2\alpha -1$. In an SSA model `m' is connected to a, b, and p as $a = 2m + 0.5$ in
the optically thick phase and $b = (p + 5 - 6m)/2$ in the optically thin phase \citep{chevalier1998}.\\

\begin{figure*}
\begin{centering}
\includegraphics[scale=0.7]{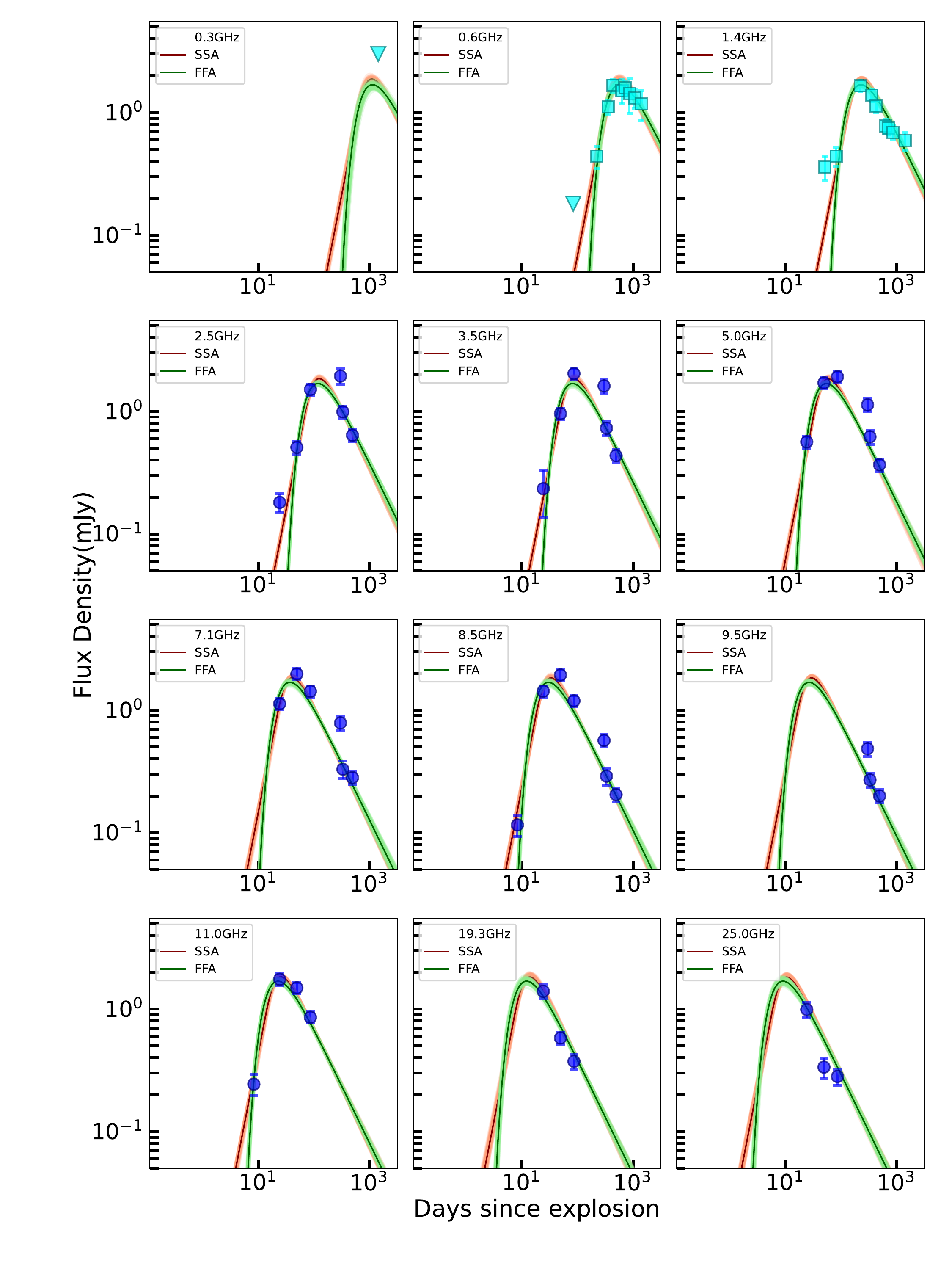}
 \caption{ \scriptsize{Radio light curves of SN\,2016gkg at frequencies $\nu =$ 0.33 $-$ 24 GHz. The solid red curves represent the best-fit SSA (equations \ref{eqn:ssa-general-flux} and \ref{eqn:ssa-optical-depth}) model and solid green curves represent best-fit FFA (equations \ref{eqn:ffa-general-flux} and \ref{eqn:ffa-optical-depth}) model (see \S \ref{sec:radio-model}). The light green (FFA) and red (SSA) lines represent 100 random draws from the MCMC posterior. The filled blue circles denote the JVLA flux density measurements and the filled cyan squares are GMRT flux density measurements. The inverted triangles denote 3$\sigma$ flux density upper limits. The fit is a 2D fit performed by including the entire data set. The age of the SN is calculated assuming the date of explosion to be 2016 Sep 20.15 UT \citep{kilpatrick2017}.}}
    \label{fig:light-curve}
    \end{centering}
\end{figure*}

\begin{figure*}
\begin{centering}
\includegraphics[scale=0.7]{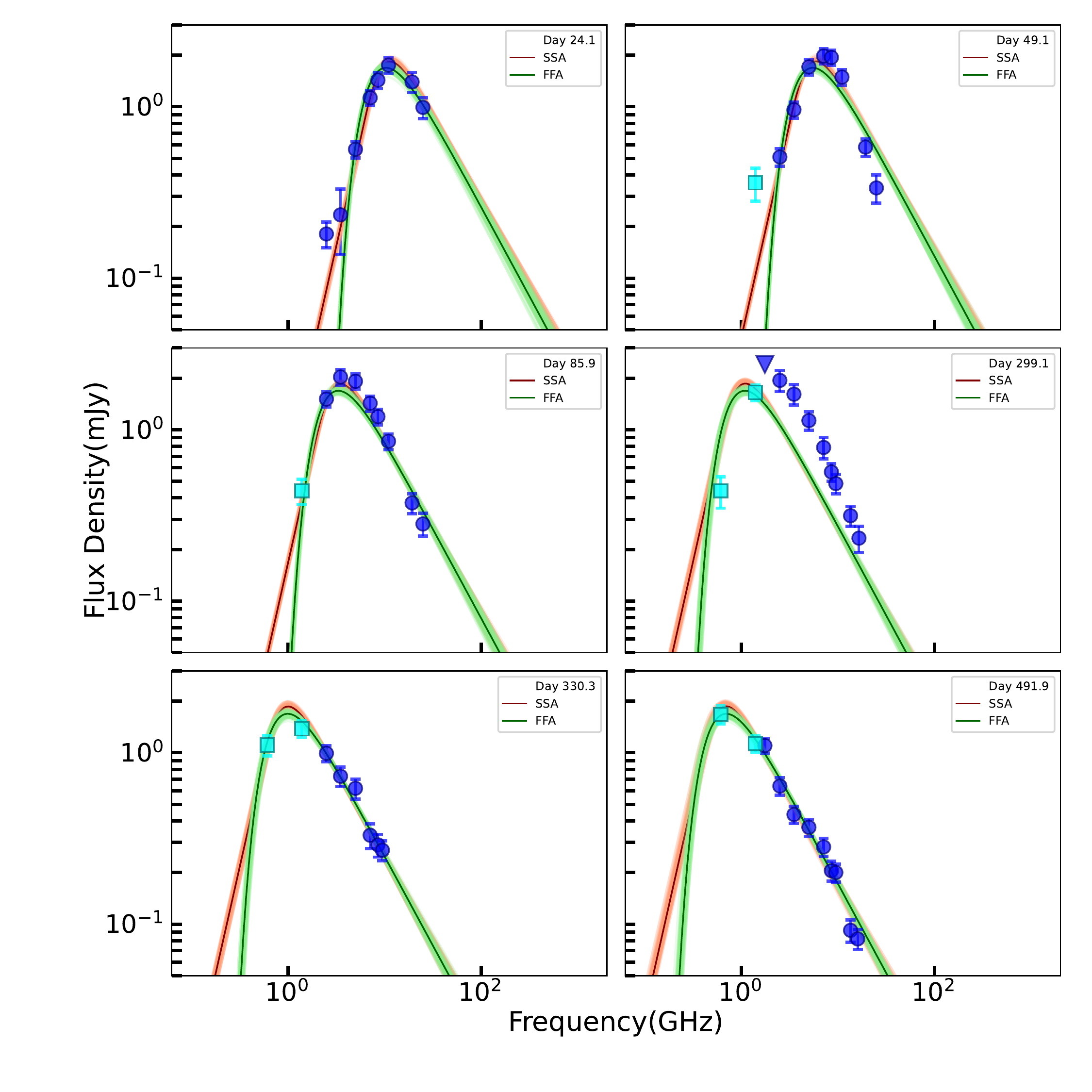}
 \caption{ \scriptsize{Radio spectra of SN\,2016gkg at  $t \sim$ 24, 49, 86, 299, 330, and 492 days post-explosion. The solid red curves represent the best-fit SSA (equations \ref{eqn:ssa-general-flux} and \ref{eqn:ssa-optical-depth}) model and solid green curves represent best-fit FFA (equations \ref{eqn:ffa-general-flux} and \ref{eqn:ffa-optical-depth}) model (see \S \ref{sec:radio-model}). The light green (FFA) and red (SSA) lines represent 100 random draws from the MCMC posterior. The filled blue circles denote the JVLA flux density measurements and the filled cyan squares are GMRT flux density measurements. The fit is a 2D fit performed by including the entire data set. The age of the SN is calculated assuming the date of explosion to be 2016 Sep 20.15 UT \citep{kilpatrick2017}.}}
    \label{fig:radio-spectra}
    \end{centering}
\end{figure*}

\begin{figure*}
\begin{centering}
\includegraphics[scale=0.38]{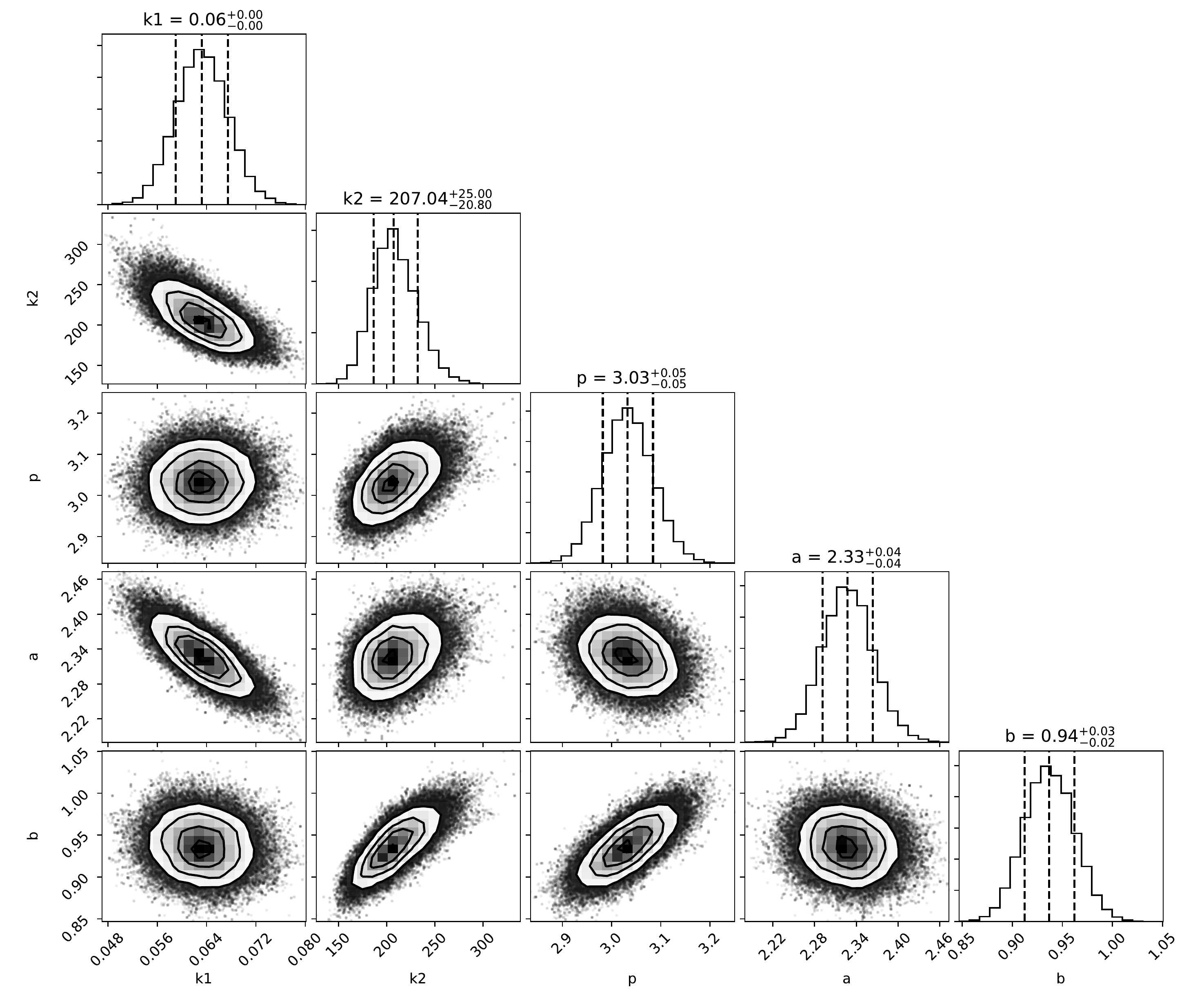} \\
\includegraphics[scale=0.38]{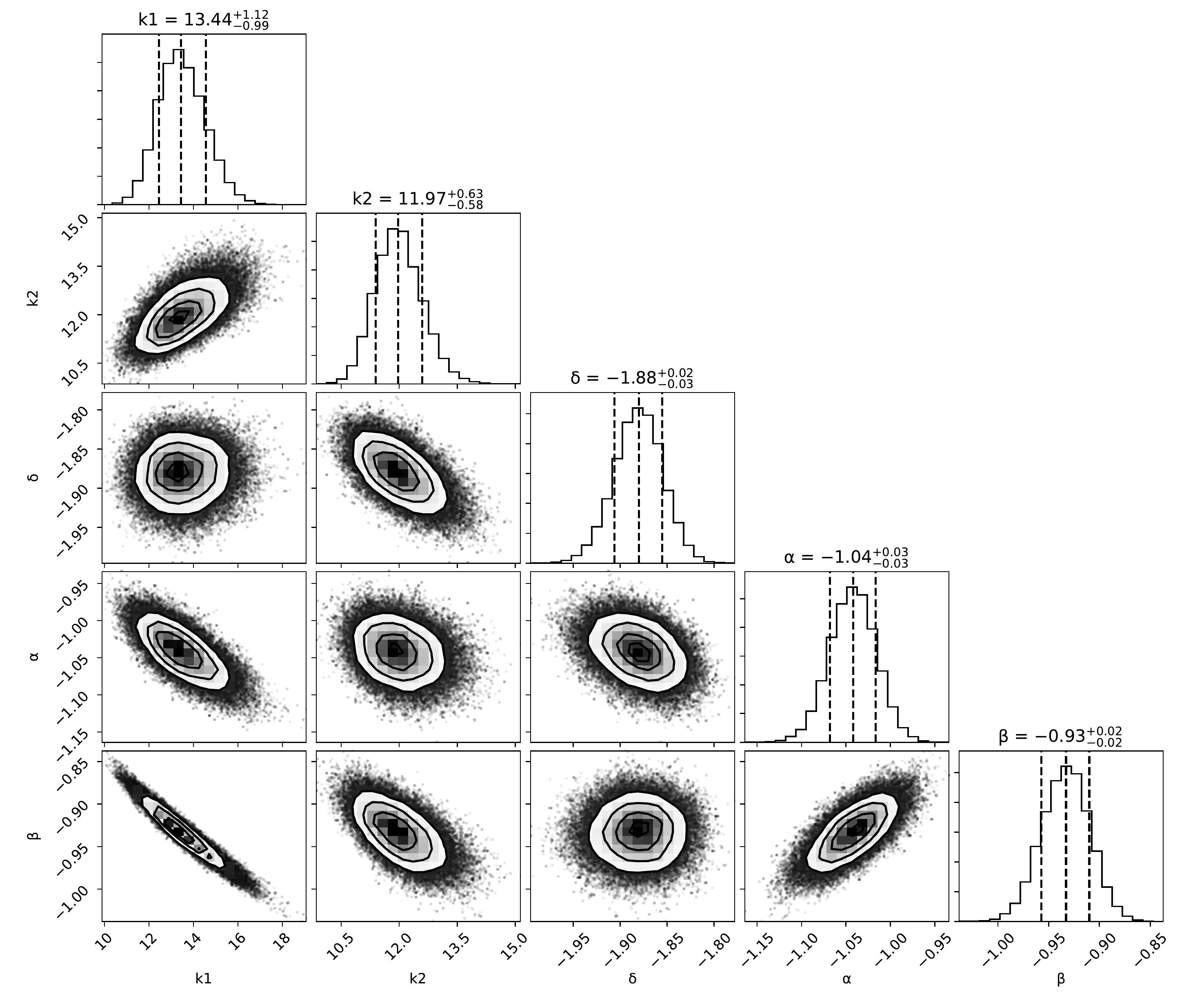}
 \caption{ \scriptsize{Corner plots that show the results of MCMC modeling of the radio data with the SSA model (equations \ref{eqn:ssa-general-flux} and \ref{eqn:ssa-optical-depth}) in top panel and FFA model (equations \ref{eqn:ffa-general-flux} and \ref{eqn:ffa-optical-depth}) in bottom panel as discussed in \S \ref{sec:radio-model}. The 16, 50, and 84 percentiles are marked in each of the contours and histograms.}}
    \label{fig:corner-ssa}
    \end{centering}
\end{figure*}

We perform a two variable fit to the entire data to find the best parameter fit to FFA model using equations \ref{eqn:ffa-general-flux} and \ref{eqn:ffa-optical-depth} and SSA model using equations \ref{eqn:ssa-general-flux} and \ref{eqn:ssa-optical-depth}. We execute the fit adopting the Markov chain Monte Carlo (MCMC) method using python package \textit{emcee} \citep{Foreman-Mackey2013}. We choose 32 walkers and 5000 steps to explore the parameter space to get the best-fit values (68\% confidence interval). We estimate the goodness of fit using the reduced $\chi^2_{\mu}$ test. We allow the parameters $K_{1}$, $K_{2}$, $\alpha$, $\beta$, and $\delta$ to vary freely in the FFA model, and $K_{1}$, $K_{2}$, $a$, $b$, and $p$ to vary freely in the SSA model. The best fit values of these parameters are listed in Table \ref{tab:2dmodel}. The best-fit modeled curves along with the observed data points are shown in Figures \ref{fig:light-curve} and \ref{fig:radio-spectra}. The corner plots are presented in Figure \ref{fig:corner-ssa}.

From the best-fit modeled curves and reduced $\chi_{\mu}^{2}$ values, the SSA model ($\chi_{\mu}^{2} =$ 3.2) seems like a better representation of the observed data compared to FFA model ($\chi_{\mu}^{2} =$ 4.4). However, there are deviations from SSA model predictions, particularly at $t \sim$ 299 days. The model underestimates the flux densities at multiple frequencies (see Figure \ref{fig:light-curve}). This effect is clear in the spectrum of day 299 where all the flux densities in the optically thin regime are systematically above the modeled curve (see Figure \ref{fig:radio-spectra}). We discuss this effect in the context of a CSM density enhancement in \S \ref{sec:non-uniform csm}. 


The best-fit value of shock deceleration parameter, $m \sim 0.6$ ($m = -\delta/3$) in FFA model. This is fairly low compared to the typical $m$ values seen in SNe\,IIb and would imply a highly decelerating shock wave which is unphysical at this early stages of evolution. The value of $m \sim$ 1 from SSA model [$m = (p+5-2b)/6$], indicative of a non-decelerating blast wave. Thus we infer SSA model to be a better representation of the data over FFA model due to lower $\chi_{\mu}^{2}$ values and the unrealistic $m$ value implied by the FFA model.

The low-frequency flux measurements at earlier epochs (t $\sim$ 24 and 49 days) are above the model predictions (see Figure \ref{fig:radio-spectra}). The spectral indices of the flux densities between 0.61 and 1.39 GHz are $0.76\pm0.45$ and $0.59\pm0.25$ at $t \sim$ 24 and 49 days, respectively. These values are flatter than the expected spectral indices ($\alpha =$ 2.5) in a standard SSA model. This can be attributed to the inhomogeneities in the magnetic fields and/or relativistic electron distribution in the
emitting region \citep{bjornsson2017,chandra2019,ho2019,Nayana2021}.

\begin{deluxetable}{c  c}
\centering
\tablecaption{Best fit parameters for FFA and SSA models.  \label{tab:2dmodel}}
\setlength{\tabcolsep}{20pt}
\tablecolumns{2}
\tablenum{3}
\tablewidth{0pt}
\tablehead{
\colhead{FFA}   & \colhead{SSA}
}
\startdata
$K_1 = 13.44_{-0.99}^{1.12}$     &
$K_1 = 0.06_{-0.00}^{0.00}$\\
$K_2 = 11.97_{-0.58}^{0.63}$     & 
$K_2 = 207.04_{-20.80}^{25.00}$\\
$\alpha = -1.04_{-0.03}^{0.03}$  & 
$a = 2.33_{-0.04}^{0.04}$ \\
$\beta = -0.93_{-0.02}^{0.02}$   & 
$b = 0.94_{-0.02}^{0.02}$ \\
$\delta = -1.88_{-0.03}^{0.02}$  & 
$p = 3.03_{-0.05}^{0.05}$\\
\hline
$\chi^2 = 4.4$                  & 
$\chi^2 = 3.2$\\
\enddata
\end{deluxetable}

\subsection{Single epoch spectral analysis}
\begin{figure}
\begin{centering}
\includegraphics[scale=0.43]{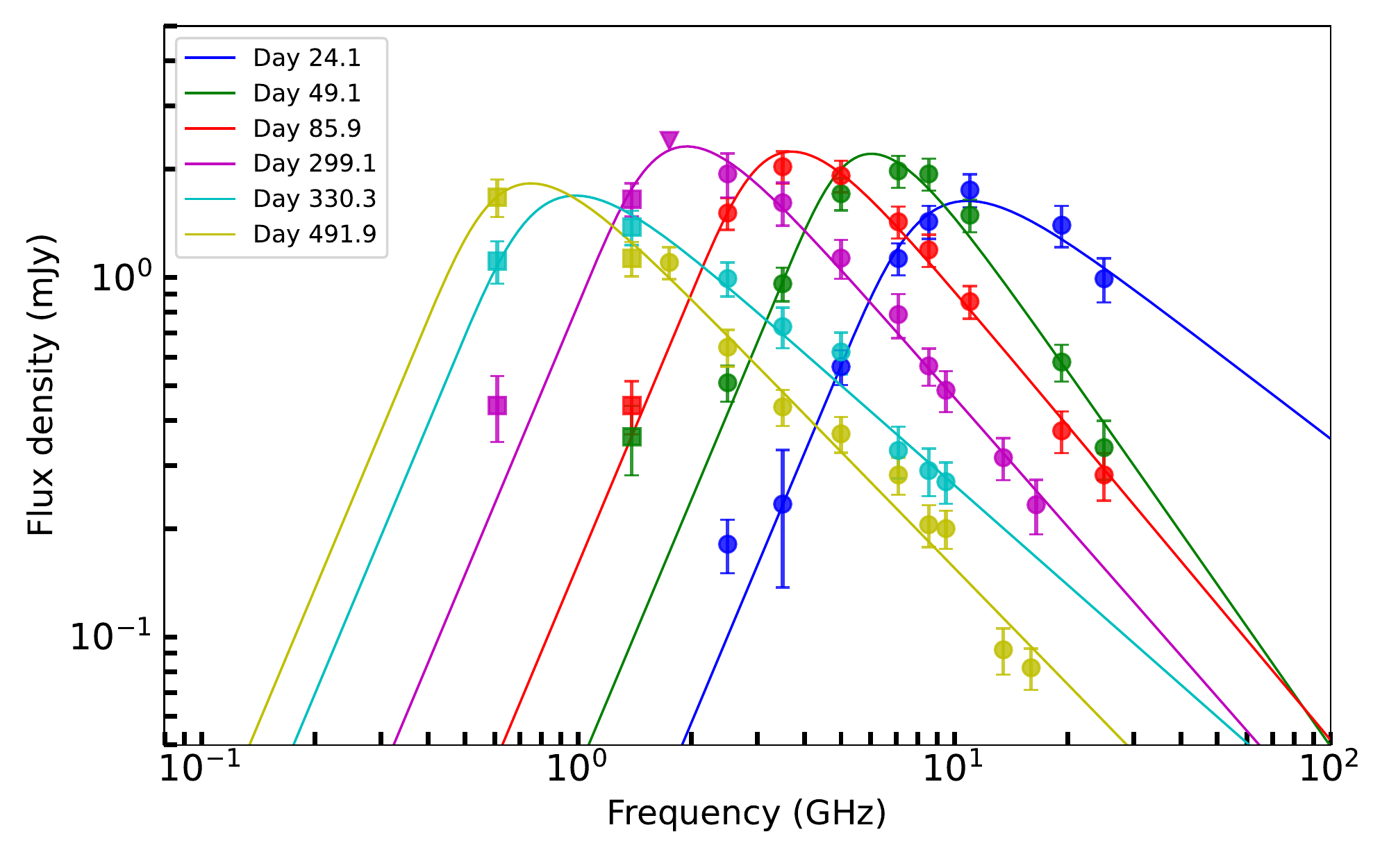} 
 \caption{ \scriptsize{Radio spectra of SN\,2016gkg at $t \sim$ 24, 49, 86, 299, 330, and 492 days post-explosion. The filled circles denote the JVLA flux density measurements and the filled squares are near-simultaneous flux density measurements from the GMRT. The solid curves are the best-fit SSA models at each epoch.}}
    \label{fig:single-epoch-deceleration}
    \end{centering}
\end{figure}

To further investigate the time evolution of shock parameters, we model each of the single epoch spectra adopting the SSA model. The functional form of the SSA spectrum can be parametrized as \citep{soderberg2006}
\begin{equation}
F_{\nu} = 1.582\, F_{p} \left( \frac{\nu}{\nu_{p}} \right)^{5/2} \left( 1-\rm exp \left[- \left( \frac{\nu}{\nu_{p}} \right)^{-(5-2\alpha)/2} \right] \right)
\end{equation}
We model single epoch spectra by allowing peak frequency ($\nu_p$), peak flux density ($F_p$), and $\alpha$ to vary freely and independently. The spectra are well fitted by an SSA model with $\chi_{\mu}^{2}$ values of 0.7 $-$ 3.3 (see Figure \ref{fig:single-epoch-deceleration} and Table \ref{tab:blast_parameters}). We attempted modeling the data with FFA model as well and the fits resulted in higher $\chi_{\nu}^{2}$ values.

The best fit $F_p$, $\nu_p$, and $\alpha$ for SSA model are given in Table \ref{tab:blast_parameters}. The quoted errors are 1$\sigma$ errors (68\% confidence interval). The temporal evolution of $\nu_{\rm p}$ and $F_{\rm p}$ are such that $\nu_{\rm p} \propto t^{-0.9}$ and $F_{\rm p} \propto t^{-0.02}$, respectively, consistent with SSA model \citep{chevalier1998}.  We obtain the average spectral index over six epochs as $\alpha \approx -1$.  The power-law index of electrons determined from the optically thin spectral index ($\alpha$) is $p=3$ ($\alpha = -(p-1)/2$). \\ \\

\subsection{Blast-wave parameters}
The shock radius ($R$) and magnetic fileds ($B$) can be estimated from $\nu_{\rm p}$ and $F_{\rm p}$ at each epoch \citep{chevalier1998}. For $p=3$, the shock radius is given by 
\begin{equation}
\begin{split}\label{eqn:radius}
R = 8.8 \times 10^{15} f_{\rm eB}^{-1/19} \left( \frac{f}{0.5} \right) ^{-1/19} \left( \frac{F_{p}}{\rm Jy} \right)  ^{9/19}\\ \left( \frac{D}{\rm Mpc} \right) ^{18/19} \left( \frac{\nu_{p}}{5\, \rm GHz} \right) ^{-1} \rm cm
\end{split}
\end{equation}

The post-shock magnetic field is given by

\begin{equation}
\begin{split}\label{eqn:B}
B = 0.58 f_{\rm eB}^{-4/19} \left( \frac{f}{0.5} \right) ^{-4/19} \left( \frac{F_{p}}{\rm Jy} \right)  ^{-2/19}\\ \left( \frac{D}{\rm Mpc} \right) ^{-4/19} \left( \frac{\nu_{p}}{5\, \rm GHz} \right) \rm G
\end{split}
\end{equation}
Here, $f_{\rm eB}$ denotes the ratio of the fraction of shock energy in relativistic electrons ($\epsilon_e$) to that in the magnetic fields ($\epsilon_B$). We assume the equipartition of energy between relativistic electrons and magnetic fields and hence use $f_{\rm eB}=1$.  $f$ is the volume filling factor of the synchrotron emitting region, taken to be 0.5 \citep{chevalier1998}. $D$ is the distance to the SN in Mpc. The mean velocity of the shock at any epoch is $V$ $\sim$ $\frac{R}{t}$.

The mass-loss rates can be deduced using the magnetic field scaling relation \citep{chevalier1998}.
\begin{equation}\label{eqn:mass-loss}
\dot{M} = \frac{B^{2} R^{2} v_{w}}{2 \epsilon_{B} V^{2}}
\end{equation} 

We deduce the shock radius ($R$) and magnetic fileds ($B$) at multiple epochs using the best fit $\nu_{\rm p}$ and $F_{\rm p}$ values at $t \sim$ 24, 49, 86, 299, 330, and 492 days using equations \ref{eqn:radius} and  \ref{eqn:B}. The physical parameters are presented in Table \ref{tab:blast_parameters}. The shock wave expands from $R \sim$ 0.5 $\times$ 10$^{16}$ cm to $R \sim$ 7.3 $\times$ 10$^{16}$ cm during $t \sim$ 24 to 492 days. The temporal evolution of shock radius can be described by an index $m =$ 0.80 $\pm$ 0.08, indicating a deccelerating blast wave (see Fig \ref{fig:deceleration-index}). The shock is slightly slowing down at $t \sim$ 299 days and later evolve consistent with the temporal evolution as described by the previous phases ($t <$ 299 days). The temporal index for post-shock magnetic field is found to be $\alpha_{\rm B} =$ $-$0.80 $\pm$ 0.09 ($B\propto t^{\alpha_{\rm B}}$).

The mass-loss rate of the progenitor star at different epochs are estimated using equation \ref{eqn:mass-loss}. We assume a wind velocity $v_{\rm w} \sim$ 200 km\,s$^{-1}$ and $\epsilon_{\rm B} = 0.33$ in the calculation. The mass-loss rates are $\dot{M} \sim$ (2.2 $-$ 5.0) $\times$ 10$^{-6}$ $M_{\odot}$\,yr$^{-1}$ during 15 $-$ 115 years before explosion (see Table \ref{tab:blast_parameters}). Besides, we note that the mass-loss rate derived from the shock parameters at $t \sim$ 299 days is relatively high, $\dot{M} \sim$ 12.6 $\times$ 10$^{-6}$ $M_{\odot}$\,yr$^{-1}$, indicating a higher mass-loss at $\sim$ 48 years prior explosion. Considering the uncertainty due to the low temporal cadence of the radio observations, the timing of the enhanced mass-loss event could be between 25$-$87 years prior explosion. Hydrodynamic wave driven outbursts during the late nuclear burning stage can create density enhancements in the CSM. However, these outbursts happen at 1-2 years prior explosion in the case of SNe\,IIb \citep{fuller2018}, which does not match with the timescales of enhanced mass-loss in SN\,2016gkg.

The $\dot{M}$ estimates will considerably vary depending on the choice of wind velocity. We choose $v_{\rm w} \sim$ 200 km\,s$^{-1}$ based on the physical properties of the progenitor star ($M \sim$ 10 $M_{\odot}$, $R_{*} \sim$ 70 $R_{\odot}$, $T_{\rm eff} \sim$ 10800 K) derived from pre-explosion imaging analysis \citep{kilpatrick2021}. The escape velocity of a 10 $M_{\odot}$ star of radius 70 $R_{\odot}$ is $\sim$ 200 km\,s$^{-1}$ and the range of wind velocity of a star of $T_{\rm eff} \sim$ 10800 K is $v_{\rm w}$ $\sim$ 100$-$300 km\,s$^{-1}$ \citep{drout2009,smith2014,yoon2017}.

\begin{deluxetable*}{ccccccccc}
\centering
\renewcommand{\arraystretch}{1.5}
\tablecaption{Blast wave paramters of SN\,2016gkg  \label{tab:blast_parameters}}
\tablecolumns{5}
\tablenum{4}
\tablewidth{0pt}
\tablehead{
\colhead{Age$^{\rm a}$}   & $F_{\rm p}$ & $\nu_{\rm p}$ & $\alpha$ & $\chi_{\mu}^{2}$ &  \colhead{$R$} & \colhead{$B$}  & \colhead{$V_{\rm s}$}     & \colhead{$\dot{M}$}  \vspace{-0.2cm}\\
\colhead{(Day)}   & (mJy) & (GHz) & - & - & \colhead{($\times$10$^{15}$cm)}    & \colhead{(Gauss)}   & \colhead{$(\times$10$^{4}$km\,s$^{-1})$}   &   \colhead{($\times$10$^{-6}$ $M_\odot$\,yr$^{-1}$)}  
}
\startdata
24      & $1.55_{-0.16}^{0.18}$ 
        & $8.96_{-0.65}^{0.68}$
        & $-0.80_{-0.22}^{0.20}$ 
        & 1.72
        & $5.09_{-0.66}^{0.68}$ 
        & $1.03_{-0.08}^{0.08}$ 
        & $2.46_{-0.32}^{0.33}$
        & $2.20_{-0.86}^{0.90}$ \\
49      & $2.20_{-0.14}^{0.14}$ 
        & $5.82_{-0.25}^{0.26}$ 
        & $-1.49_{-0.13}^{0.13}$ 
        & 3.30 
        & $9.26_{-1.00}^{1.01}$ 
        & $0.65_{-0.03}^{0.03}$ 
        & $2.19_{-0.24}^{0.24}$
        & $3.58_{-1.16}^{1.16}$ \\
86      & $2.22_{-0.14}^{0.14}$ 
        & $3.43_{-0.23}^{0.28}$ 
        & $-1.25_{-0.09}^{0.09}$ 
        & 0.66 
        & $15.77_{-1.90}^{2.02}$ 
        & $0.38_{-0.03}^{0.03}$
        & $2.13_{-0.26}^{0.27}$
        & $3.82_{-1.40}^{1.52}$ \\
299     & $2.28_{-0.19}^{0.19}$
        & $1.79_{-0.13}^{0.12}$
        & $-1.19_{-0.09}^{0.09}$
        & 0.98
        & $30.59_{-3.81}^{3.70}$
        & $0.20_{-0.01}^{0.01}$ 
        & $1.18_{-0.15}^{0.14}$ 
        & $12.56_{-4.80}^{4.62}$ \\
330     & $1.62_{-0.14}^{0.15}$ 
        & $0.85_{-0.06}^{0.06}$ 
        & $-0.92_{-0.07}^{0.07}$ 
        & 0.72 
        & $55.08_{-6.93}^{6.88}$
        & $0.10_{-0.01}^{0.01}$ 
        & $1.93_{-0.24}^{0.24}$ 
        & $3.68_{-1.42}^{1.42}$ \\
492     & $1.79_{-0.14}^{0.15}$ 
        & $0.67_{-0.06}^{0.06}$ 
        & $-1.07_{-0.05}^{0.05}$ 
        & 1.63
        & $72.57_{-10.00}^{9.51}$ 
        & $0.08_{-0.01}^{0.01}$
        & $1.71_{-0.24}^{0.22}$
        & $5.02_{-2.20}^{2.06}$ \\
\enddata
\tablenotetext{a}{The age is calculated using 2016 Sep 20.15 (UT) as the date of explosion \citep{kilpatrick2017}.}
\tablenotetext{}{$F_{\rm p}$, $\nu_{\rm p}$ and $\alpha$ denote the peak flux density, peak frequency and optically thin spectral index in the best fit SSA model at each epochs. $R$, $B$, $V$, and $\dot{M}$ denote shock radius, magnetic field, mass-loss rate and mean shock velocity, respectively.}
\end{deluxetable*}

\section{Non-uniforn density of the CSM}
\label{sec:non-uniform csm}
The overall evolution of radio light curves and spectra of SN\,2016gkg is best modeled by a self-absorbed synchrotron emission that arises due to the interaction of SN shock with the CSM created by a uniform mass-loss from the progenitor. However, there are some deviations from the smooth light curve/spectra evolution as prescribed by the standard model. There is a fractional increase by a factor of $\sim$ 2 in flux densities (in the optically thin phase) at different frequencies on day 299 above the model prediction (see Fig \ref{fig:light-curve} and \ref{fig:radio-spectra}). This abrupt rise in flux densities could be due to the interaction of the forward shock with density enhancements in the CSM at a radius $R \sim$ 3.1 $\times$ 10$^{16}$ cm. 
These density fluctuations could be either due to the non-uniform mass-loss rate of a single star progenitor via stellar winds and/or due to the mass stripping by a binary companion \citep{soderberg2006}. There are several observational pieces of evidence from supernova remnants and massive stars that support complex mass-loss events happening towards the end stages of stellar evolution \citep{soderberg2006}. In the case of a binary scenario, the strength and position of CSM density enhancement will be influenced by the binary parameters \citep{podsiadlowski1992}. A build-up of CSM material can happen at particular spatial scales due to the modulation of progenitor stellar wind depending on the orbital period of the binary companion \citep{weiler1992} and the eccentricity of the binary orbit. A binary scenario with a period of 4000 years has been attributed to the periodic modulations in the radio light curves of SN\,1979C \citep{weiler1992,montes2000}.

\begin{figure}
\begin{centering}
\includegraphics[scale=0.45]{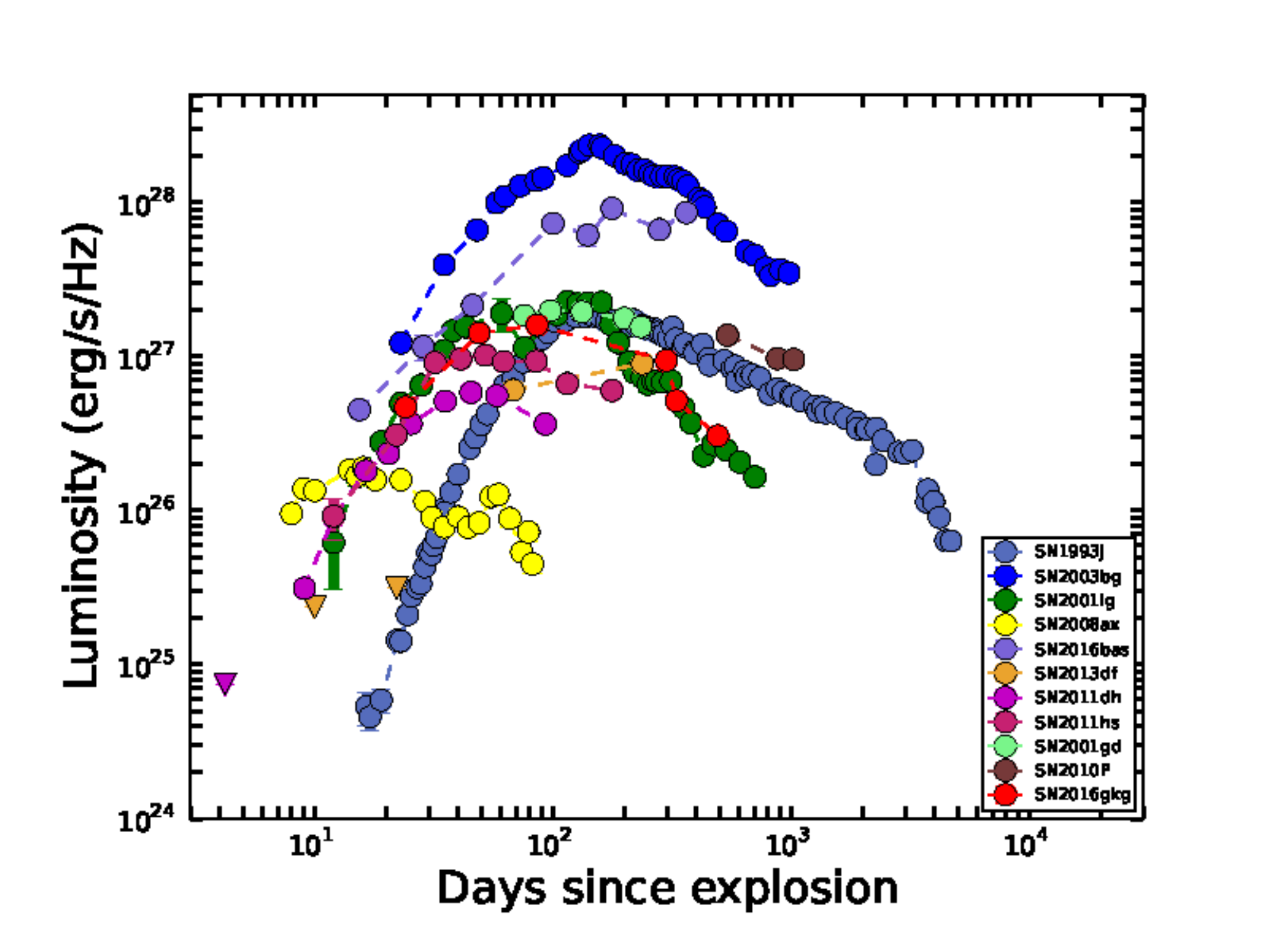}
 \caption{5 GHz radio light curves of well-observed SNe\,IIb. Flux measurements are taken from \cite{weiler2007} for SN\,1993J, \cite{soderberg2006} for SN\,2003bg, \cite{ryder2004} for SN\,2001ig, \cite{roming2009} for SN\,2008ax, \cite{bietenholz2021} for SN\,2016bas, \cite{kamble2016} for SN\,2013df, \cite{krauss2012} for SN\,2011dh, \cite{bufano2014} for SN\,2011hs, \cite{stockdale2007} for SN\,2001gd, \cite{canizales2014} for SN\,2010P, and This work for SN\,2016gkg.}
    \label{fig:radio-5ghzlum-IIbs}
    \end{centering}
\end{figure}

Multiple episodes of mass-loss events have been attributed to periodic light curve bumps of modest (factor of $\sim$ 2) flux density fluctuations in SN\,2003bg \citep{soderberg2006} and SN\,2001ig \citep{ryder2004}. Both these SNe showed variations in the light curve during a period $t =$120$-$300 days (see Fig \ref{fig:radio-5ghzlum-IIbs}), indicating a radial distance of 4$\times$10$^{16}$ cm to 8 $\times$10$^{16}$ cm from the explosion center, similar to that of SN\,2016gkg. We note that the temporal cadence of the follow-up observations of these SNe are good enough to map the periodic bumps in their light curves (see Fig \ref{fig:radio-5ghzlum-IIbs}). 


In the case of SN\,2016gkg, we see flux density enhancement at $t \sim$ 299 days, that correspond to a stellar evolution phase of $\sim$ 48 years prior explosion for $v_{\rm w} \sim$ 200 km\,s$^{-1}$. \cite{kilpatrick2017} suggested a binary progenitor model for SN\,2016gkg where the initial period of the binary orbit is $\sim$ 1000 days (2.7 years). This periodicity will be seen as another flux density enhancement at $t \sim$ 313 days for the derived shock velocities. This epoch is not sampled in the observations. Thus even if there is a binary companion and related periodicity in the density distribution of CSM, the radio data do not have a temporal cadence to probe those fluctuations, and we cannot rule out the binary scenario. 

\begin{figure}
\begin{centering}
\includegraphics[scale=0.6]{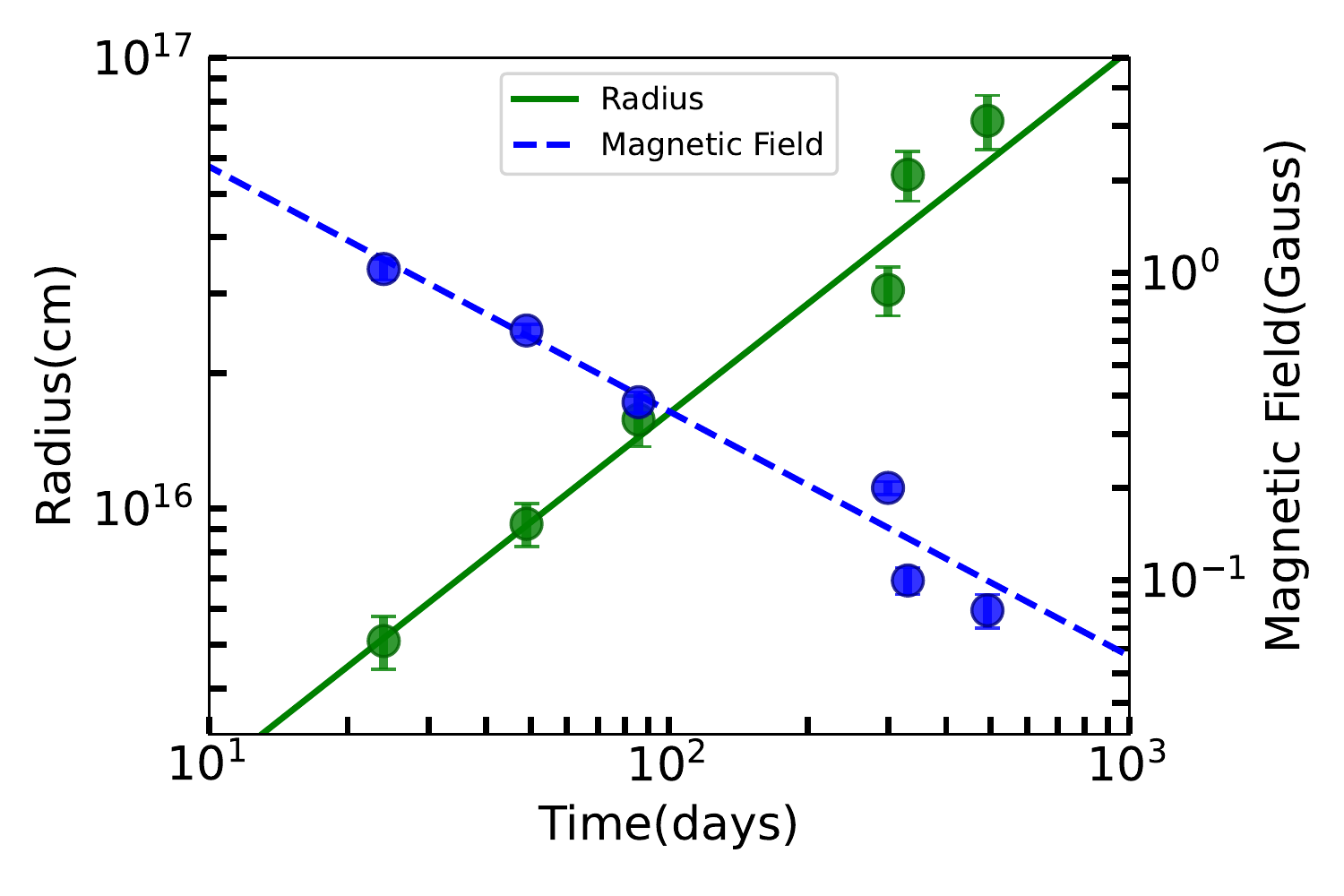}
\caption{ \scriptsize{The shock radii (green filled circles) and magnetic {\bf fields} (blue filled circles) at $t \sim$ 24, 49, 86, 299, 330, and 492 days post-explosion. The green solid line denotes a power-law fit to the shock radii; $R \sim$ $t^{0.80 \pm 0.08}$ and blue solid line denotes a power-law fit to the magnetic field; $B \sim$ $t^{-0.80 \pm 0.09}$.}}
\label{fig:deceleration-index}
\end{centering}
\end{figure}

The radio luminosity is related to the density of CSM as $L \propto$ $\rho_{\rm CSM}^{(p-7+12m)/4}$, where $\rho_{\rm CSM} \propto (\dot{M}/v_{\rm w})$ \citep{ryder2004}. In the case of SN\,2016gkg, the best fit value of $p =3$ and $m = 0.8$ indicates that a factor of 2 increase in radio flux density indicates $\sim$70\% increase in the CSM density. The effect of density enhancement is reflected in the evolution of shock radii and magnetic field as well (see Fig \ref{fig:deceleration-index}). The magnetic field is slightly enhanced compared to its expected regular temporal evolution and there is a slight change in the expansion of shock wave at the same time. The magnetic field is enhanced by a factor of 1.6 in comparison to the extrapolation of its evolution in the previous phases. The additional thermal energy produced due to these density enhancements increases the $B$. A similar increase in $B$ value by a factor of 1.3 is seen in SN\,2003bg during its first light curve bump \citep{soderberg2006}. 
To summarize, there is a close resemblance between SN\,2016gkg, and SN\,2003bg, and SN\,2001ig in terms of the timescale and strength of flux density enhancement in late-time radio light curves.

\section{Discussion}
\label{sec:discussion}
The shock velocities of SN\,2016gkg derived from SSA modeling is $v \sim$ 24600 km\,s$^{-1}$ ($v \sim$ 0.1 c) at $t \sim$ 24 days. The velocities estimated from optical lines at  $t \sim$ 21 days is $v \sim$ 12,200 km\,s$^{-1}$ \citep{tartaglia2017}. Thus the shock wave is traveling with a velocity a factor of 2 faster than the material in the photosphere. The SSA-derived shock velocity is greater than the velocities from optical lines and indicates that FFA is not contributing much in defining the peak of the light curve/spectra. We also note that the FFA models were resulting in higher $\chi_{\mu}^{2}$ values while modeling. The radio data being inconsistent with the FFA model and relatively higher shock velocity of $v \sim$ 0.1 c are indicative of a compact progenitor star with faster stellar winds.

The temporal evolution of shock wave radius is best fitted by a power law $R \propto$ $t^{0.80 \pm 0.08}$. The $m$ value will be 1, for a non-deccelerating blast wave and the derived $m$ value indicates a decelerating blast-wave. The temporal index of post-shock magnetic field is found to be $\alpha_{\rm B}$ = $-$0.8$\pm$0.1 ($B\propto t^{\alpha_{\rm B}}$). $\alpha_{\rm B}$ can be connected to the CSM density as $\alpha_{\rm B} = [m(2-s)/2]-1$, which gives $s = 1.5$. Thus the CSM density is slightly flatter than the one created by a steady stellar wind. The derived values of $m$ and $s$ indicate an ejecta density profile of $\rho_{\rm ejecta} \propto r^{-9}$ ($m = n-3/n-s$), consistent with low-mass compact progenitor \citep{chevalier1998}. Assuming $\epsilon_{\rm B} =$ 0.33 and $v_{\rm w} =$ 200 km\,s$^{-1}$, the mass-loss rate of the progenitor is in the range $\dot{M} \sim$ (2.2--5.0) $\times$ 10$^{-6}$ $M_{\odot}\,yr^{-1}$, at $t \sim$ 24, 49, 86, 330, and 492 days. The mass-loss rate corresponding to $t \sim$ 299 days is $\dot{M} \sim$  12.6 $\times$ 10$^{-6}$ $M_{\odot}\,yr^{-1}$, a factor of three higher than the $\dot{M}$ values at other epochs. This is suggestive of an enhanced phase of mass-loss at $\sim$ 48 years prior explosion for the assumed wind speed as discussed in \S \ref{sec:non-uniform csm}.  

\begin{figure}
\begin{centering}
\includegraphics[scale=0.43]{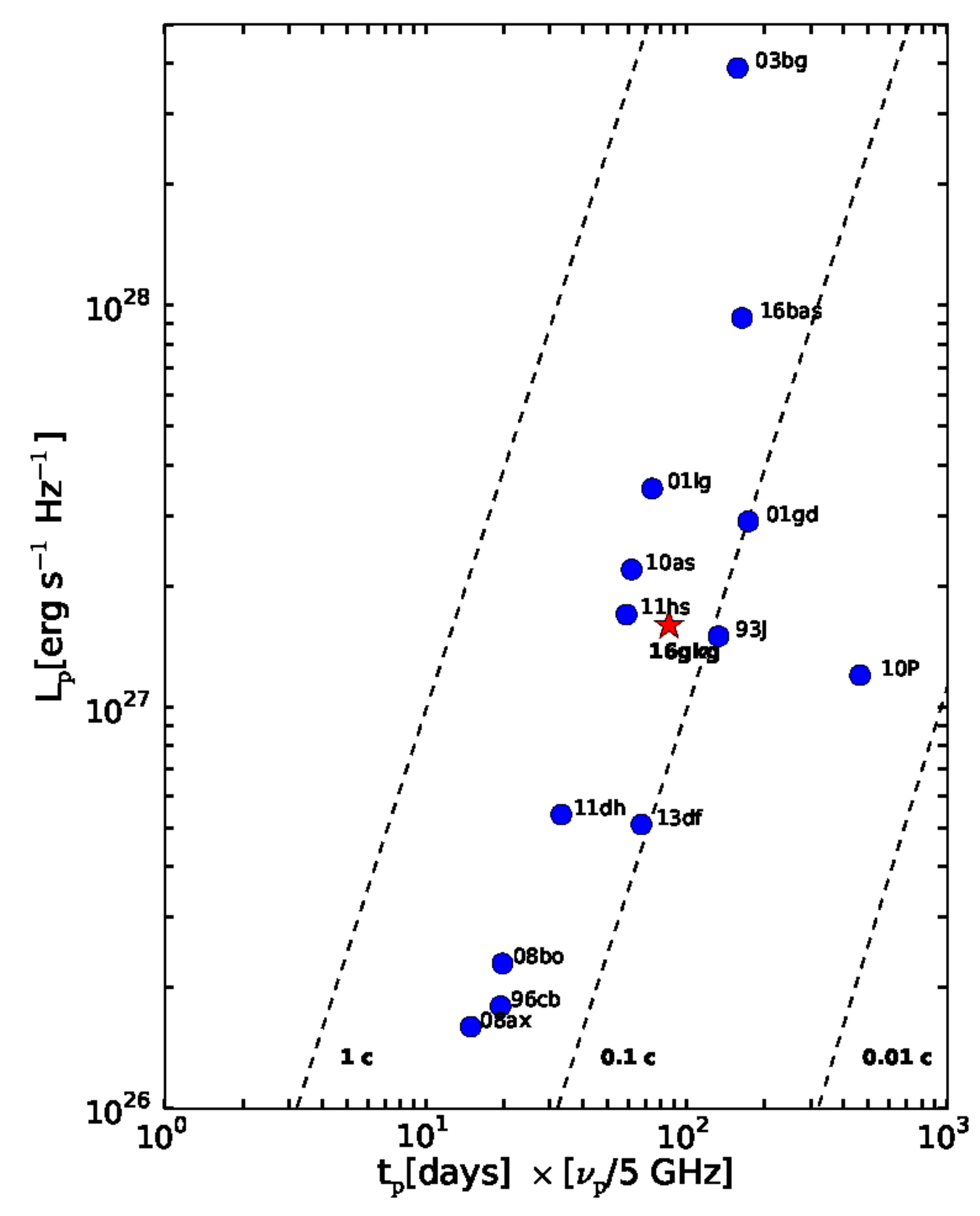}
 \caption{The peak radio spectral luminosities versus time of peak of the light curves are plotted for all well-observed SNe\,IIb in the literature. Each SNe is denoted by the last two digits of the year of explosion and last two letters of the name. The dotted lines represent the mean velocity of the radio emitting shell if SSA is the dominant absorption process that defines the peak of the light curve \citep{chevalier1998} and for $p =$ 3.  }
    \label{fig:radio-sneIIb-comparison}
    \end{centering}
\end{figure}

\begin{deluxetable*}{ccccccccc} 
\centering
\renewcommand{\arraystretch}{1.5}
\tablecaption{Properties of well-studied type IIb supernovae  \label{tab:discussion}}
\tablecolumns{8}
\tablenum{5}
\tablewidth{0pt}
\tablehead{
\colhead{SN}        &   \colhead{Distance}  & \colhead{$\nu_p$}     & \colhead{$t_p$} & \colhead{$F_p$}     & \colhead{$L_{\rm p}$} & \colhead{$\dot{M}$}   & \colhead{References}\\
\colhead{-}       & \colhead{(Mpc)} & \colhead{(GHz)}     & \colhead{(days)}      & \colhead{(mJy)} & \colhead{(erg\,s$^{-1}$\,Hz$^{-1}$)} & \colhead{($M\odot yr^{-1}$)}        & \colhead{-}
}
\startdata
SN\,1993J  & 3.6  & 5    & 133   & 96.9 &  1.5 $\times$ 10$^{27}$ & (2$-$6) $\times$10$^{-5}$ & 1 \\
SN\,1996cb & 9.1  & 5    & 19.4  & 1.8  &  1.8 $\times$ 10$^{26}$ & -- & 2\\
SN\,2001gd & 17.5 & 5    & 173    & 8.0    & 2.9 $\times$ 10$^{27}$ & 3$\times$10$^{-5}$   & 3\\
SN\,2001ig & 11.5 & 5    & 74   & 22   & 3.5 $\times$ 10$^{27}$  & 8.6$\times$10$^{-5}$  & 4 \\
SN\,2003bg & 19.6 & 22.5 & 35    & 85   & 3.9 $\times$ 10$^{28}$ & 6.1 $\times$ 10$^{-5}$   & 5\\
SN\,2008ax & 6.2  & 4.86    & 15.34    & 3.54    & 1.6 $\times$ 10$^{26}$ & (1-6) $\times$ 10$^{-6}$ & 6 \\
SN\,2008bo & 19.1 & 8.5  & 11.6  & 0.52 & 2.3 $\times$ 10$^{26}$ & -  & 7 \\
SN\,2010P  & 44.8 & 5    & 464   & 0.52 & 1.2 $\times$ 10$^{27}$ & (3.0$-$5.1)$\times$10$^{-5}$& 8 \\
SN\,2010as & 27.4 & 9    & 34.3  & 2.43 & 2.2 $\times$10$^{27}$ & - & 7 \\
SN\,2011dh & 7.9  & 4.7  & 35.3  & 7.3  & 5.4 $\times$ 10$^{26}$  & 6 $\times$10$^{-5}$ & 9 \\
SN\,2011hs & 26.4 & 5    & 59    & 2.0  & 1.7 $\times$ 10$^{27}$ & 2 $\times$ 10$^{-5}$  & 10 \\
SN\,2013df & 16.6 & 5    & 67.3  & 1.55 & 5.1 $\times$10$^{26}$ & 8 $\times$ 10$^{-5}$ & 11\\
SN\,2016bas & 42.4 & 5.5 & 148.8 & 4.33 & 9.3 $\times$10$^{27}$ & - & 7 \\
{\bf SN\,2016gkg} & {\bf 26.4} & {\bf 5}   & {\bf 85.86} & {\bf 1.92} & {\bf 1.6 $\times$10$^{27}$ } & {\bf 3.8$\times$10$^{-6}$} & 12 \\ 
\enddata
\tablerefs{(1) \cite{fransson1996}, (2)  \cite{weiler1998}, (3) \cite{stockdale2003}, (4) \cite{ryder2004}, (5) \cite{soderberg2006}, (6) \cite{roming2009}, (7) \cite{bietenholz2021}, (8) \cite{canizales2014}, (9) \cite{soderberg2012}, (10) \cite{bufano2014}, (11) \cite{kamble2016}, (12) This work }
\tablecomments{Among the listed SNe, five of them have progenitor detections from archival images. They are SN\,1993J \citep{aldering1994}, SN\,2013df \citep{vandyk2014}, SN\,2008ax \citep{folatelli2015}, SN\,2011dh \citep{vandyk2013,arcavi2011} and SN\,2016gkg \citep{kilpatrick2017,kilpatrick2021,tartaglia2017}.}
\tablecomments The listed $\dot{M}$ values are taken from the literature as cited in the last column of the table. These values are strongly dependent on the assumed wind velocities.
\end{deluxetable*}

\begin{deluxetable*}{ccc}
\centering
\tablecaption{Compilation of the radius estimates of the progenitor of SN\,2016gkg from the literature}
\label{radius}
\tablecolumns{3}
\tablewidth{0pt}
\tablehead{
\colhead{Radius ($R_{\odot}$)}    &\colhead{Method}   &\colhead{Reference}
}
\startdata
 138$^{+131}_{-103}$     &Archival HST imaging analysis\tablenotemark{a}                    &\cite{kilpatrick2017} \\
 150 $-$ 320    &Archival HST imaging analysis\tablenotemark{a}                    &\cite{tartaglia2017} \\
 $\sim$ 70      &Archival HST imaging analysis\tablenotemark{b}                    &\cite{kilpatrick2021}\\
 257$^{+389}_{-189}$     &Analytical shock cooling model\tablenotemark{c}          &\cite{kilpatrick2017}\\
 48 $-$ 124     &Analytical shock cooling model\tablenotemark{c}          &\cite{tartaglia2017}\\
 40 $-$ 150     &Analytical shock cooling model\tablenotemark{d}          &\cite{arcavi2017}\\
 180 $-$ 260    &Numerical shock cooling model                          &\cite{piro2017} \\
 $\sim$ 320 $R_{\odot}$   &Numerical shock cooling model                          &\cite{bersten2018}\\
\enddata
\tablenotetext{a}{\cite{kilpatrick2017} and \cite{tartaglia2017} used Keck and Very Large Telescope + Nasmyth adaptive optics systems, respectively to perform relative astrometry. \cite{kilpatrick2017} considered one progenitor candidate and \cite{tartaglia2017} considered two progenitor candidates. }
\tablenotetext{b}{Post-explosion HST imaging of the field containing SN\,2016gkg was used to perform relative astrometry.}
\tablenotetext{c}{\cite{kilpatrick2017} and \cite{tartaglia2017} modeled the luminosity (up to $t \sim$ 1.5 days) and temperature evolution (up to $t \sim$ 5 days), respectively using \cite{rabinak2011} model.}
\tablenotetext{d}{Analytic shock cooling models \citep{piro2015,nakar2014,sapir2017}.}
\end{deluxetable*}

\subsection{Comparison with other radio bright SNe\,IIb}
\label{subsec:comparison-all-SneIIb}
We view the radio properties of SN\,2016gkg in comparison with other radio bright SNe\,IIb events in Figure \ref{fig:radio-sneIIb-comparison}. We compile all SNe\,IIb with well sampled light curves/spectra that defines $F_{\rm p}$, $\nu_{\rm p}$ and $t_{\rm p}$ in the $L_{\rm p}$ - $t_{\rm p}$ diagram (also see Table \ref{tab:discussion}). The dotted lines indicate the mean shock velocities in an SSA scenario for $p = 3$ assuming equipartition of energy between relativistic particles and magnetic fields ($\epsilon_{\rm B}$ = $\epsilon_{\rm e}$). This plot is an updated version of a similar plot presented in \cite{chevalier2010}. The authors compiled radio properties of a sample of SNe\,IIb and divided them into two populations based on their position in the $L_{\rm p}$ - $t_{\rm p}$ diagram, one is SNe\,IIb with compact progenitors (SNe\,cIIb) and the other with extended progenitors (SNe\,eIIb). The SNe\,cIIb group consists of SN\,2008ax, SN\,2003bg and SN\,2001ig which has faster shocks, less dense CSM and compact progenitor in comparison with the SNe\,eIIb (like SN\,1993J and SN\,2001gd). SNe\,eIIb have slower shocks owing to its denser CSM from slow stellar winds of extended progenitors. 

SNe\,IIb show peak spectral luminosities that span 2 orders of magnitude, and the peak time vary over a factor of $\sim$ 40. This broad distribution in peak spectral luminosities and rise times indicates the variety in the intrinsic properties of their progenitors. There are only 5 SNe\,IIb (including SN\,2016gkg) that have progenitors identified from pre-explosion images. It is important to tie up the radio properties of these SNe with the inferences from pre-explosion imaging analysis. The progenitor of SN\,1993J is identified to be a YSG of radius $\sim$ 600 $R_{\odot}$ \citep{aldering1994} from pre-explosion images. Similarly, an extended progenitor of radius $\sim$ 545 $\pm$ 65 $R_{\odot}$ was identified as the progenitor of SN\,2013df \citep{vandyk2014}. A slightly less extended star of radius $\sim$ 200 $R_{\odot}$ was identified as the progenitor of SN\,2011dh \citep{vandyk2013}. However, the presence of a binary companion has been speculated for this SN \citep{vandyk2013}, and \cite{arcavi2011} suggested the progenitor of SN\,2011dh to be a relatively compact star from the analysis of a series of spectra and bolometric light curve. The authors also argued that the larger radius ($R \sim$ 10$^{13}$ cm) derived from pre-explosion HST images could be due to the identification of a blended source. The progenitor of SN\,2008ax was identified to be of radius $\sim$ 30$-$50 $R_{\odot}$ \citep{folatelli2015}. These estimates of progenitor radius from pre-SN image analysis is roughly consistent with the radio-derived properties. SN\,1993J and SN\,2013df to have more extended progenitors whereas SN\,2008ax and SN\,2011dh to have relatively compact progenitors with higher shock velocity (see Figure \ref{fig:radio-sneIIb-comparison}). In light of these results, one can also argue that the classification of SNe\,IIb progenitors into two categories as eSNe\,IIb and cSNe\,IIb \citep{chevalier2010} is rather simplistic and the progenitor properties could be a continuum between these two. The position of SN\,2016gkg in this diagram is among SNe\,cIIb, towards the right of SN\,2008ax. This could imply that the progenitor of SN\,2016gkg is a relatively compact progenitor with a radius slightly more than that of SN\,2008ax, i.e., 50 $R_{\odot}$ .

\subsection{Inferences on the progenitor}
Multiple pieces of evidence from radio modeling are in favor of a compact progenitor for SN\,2016gkg. The broad agreement of multi-frequency radio data with an SSA model indicates that the CSM is relatively rarer created due to faster stellar winds from a compact star. The mean shock velocities ($v \sim$ 0.1 c) derived from SSA formulation are difficult to incorporate in the framework of a shock breakout from an extended progenitor \citep{nakar2010}. 

A correlation between $\dot{M}$ values and progenitor radius of SNe\,IIb is proposed by \cite{maeda2015} and \cite{kamble2016}, where the extended progenitors experience stronger mass-loss towards their end-of-life compared to compact progenitors. The progenitor mass-loss rate of SN\,2016gkg derived from the shock parameters at $t \sim$ 24 days is $\dot{M} \sim$ 2.2 $\times$ 10$^{-6}$ $M_{\odot}$\,yr$^{-1}$, comparable to the $\dot{M}$ values derived for SN\,2008ax \citep{roming2009}. These $\dot{M}$ values are an order of magnitude lower than that of SN\,1993J and SN\,2013df (see Table \ref{tab:discussion}) which are known to have extended progenitors from direct detection efforts.  Thus the $\dot{M}$ estimates also imply a relatively compact progenitor. 

The best-fit value of electron power-law index is $p \sim$ 3, typically found for SNe\,Ibc which are presumed to have compact WR stars as progenitors. Late-time variability in the radio light curves is an important observational characteristic of SNe\,cIIb progenitors \citep[e.g., SN\,2001ig, SN\,2003bg][]{ryder2004,soderberg2006}.
All SNe\,cIIb except SN\,2011dh with well-sampled radio light curves exhibit fluctuations indicative of density modulations in the CSM. These density fluctuations could be due to the variability in the stellar winds of compact stars or due to the influence of a binary companion. The radio observations of SN\,2011dh probe a radius up to 1.5 $\times$ 10$^{16}$ cm that translates to $\sim$ 5 years prior explosion for a $v_{\rm w} \sim$ 1000 km\,s$^{-1}$, which could be shorter for any substantial wind variability \citep{krauss2012}. We see similar late-time flux density enhancement in the radio light curves of SN\,2016gkg. The position of SN\,2016gkg in the $L_{\rm p} - t_{\rm p}$ diagram is in the contour of SNe\,cIIb, indicating a progenitor radius slightly more than that of SN\,2008ax (i.e., $>$ 50 $R_{\odot}$).

The radius estimates of the progenitor from shock cooling models span a wide range $\sim$ 40 -- 646 $R_{\odot}$ \citep{bersten2018,kilpatrick2017,tartaglia2017,piro2017,arcavi2017}. One can argue that the radio derived constraints of a compact progenitor is broadly in agreement with the results from shock cooling models. The large range of radius estimates from these models will be in agreement with an extended progenitor model as well.

\cite{kilpatrick2017} and \cite{tartaglia2017} determined the progenitor radii to be $R =$138$^{+131}_{-103}$ $R_{\odot}$ and $R =$150$-$320 $R_{\odot}$, respectively, from the pre-explosion \textit{HST} imaging analysis of the field containing SN\,2016gkg. Relative astrometry was done using Keck and VLT adaptive optics system in these studies. The late-time HST imaging of the field of SN\,2016gkg using Advanced Camera for Surveys (ACS) and Wide Field Camera 3 (WFC 3) resulted in superior resolution and improved astrometric alignment between the SN and progenitor candidate \citep{kilpatrick2021}. The updated photometric analysis suggests the progenitor to be a yellow supergiant of mass 10$^{+2}_{-1}$ $M_{\odot}$ and radius $\sim$ 70 $R_{\odot}$. Thus the inferences on the progenitor star from radio analyses is in agreement with that derived from pre-explosion imaging analysis \citep{kilpatrick2021}.

\section{Conclusions}
\label{sec:conclusion}
 We present long-term ($t \sim$ 8 $-$ 1429) radio monitoring of SN\,2016gkg over a frequency range $\nu \sim$ 0.3 $-$ 24 GHz to investigate the properties of its progenitor and CSM. The inferences from our observations and modeling can be summarized as follows
\begin{itemize}
\item The radio data is best represented by a self-absorbed synchrotron emission that arises due to the interaction of an SN shock wave of $v \sim$ 0.1 c propagating into a CSM created due to the mass-loss of the progenitor star. 
\item The CSM density is found to have moderate density fluctuation at a distance $R \sim$ 3.1 $\times$ 10$^{16}$ cm, likely due to enhancement in the progenitor mass-loss rate or due to the effect of a binary companion. Assuming a stellar wind velocity $v_{\rm w} \sim$ 200 km\,s$^{-1}$, this corresponds to a stellar evolution phase $\sim$ 48 years prior explosion.
\item We estimate the average mass-loss rate to be $\dot{M} \sim$ 3.7 $\times$ 10$^{-6}$ $M_{\odot}\,yr^{-1}$ during 8 to 115 years before explosion, with a factor of $\sim$ 3 higher $\dot{M}$ at $\sim$ 48 years prior explosion. 
\item The radio data being consistent with SSA model, shock velocities of $v \sim$ 0.1 c, the position of SN\,2016gkg in the region of SNe\,cIIb in $L_{\rm p}$-$t_{\rm p}$ diagram, and late time modest variability in radio flux densities are suggestive of a compact progenitor star.

\end{itemize}

\acknowledgments
We thank the staff of the GMRT that made these observations possible. GMRT is run by the National Centre for Radio Astrophysics of the Tata Institute of Fundamental Research.
The National Radio Astronomy Observatory is a facility of the National Science Foundation operated under cooperative agreement by Associated Universities, Inc. Nayana A.J. acknowledges DST-
INSPIRE Faculty Fellowship (IFA20-PH-259) for supporting this research. P.C. acknowledges the support of the
Department of Atomic Energy, Government of India, under project no. 12-R\&D-TFR-5.02-0700.

\vspace{5mm}
\facilities{Giant Metrewave Radio Telescope, Karl J. Jansky Very Large Array}


\software{CASA \citep{mcmullin2007}, AIPS \citep{van1996}, emcee \citep{Foreman-Mackey2013}
          }



\clearpage

\end{document}